\documentclass[a4paper]{article}

\usepackage{INTERSPEECH2021, multirow}

\title{A Mixture of Expert Based Deep Neural Network for Improved ASR}
\name{Vishwanath Pratap  Singh$^1$, Shakti P. Rath$^2$, and Abhishek Pandey$^1$}
\address{
	$^1$Samsung R\&D Institute India - Bangalore\\
	$^2$Reverie Language Technologies, India}
\email{vp.singh@samsung.com, shakti.rath@reverieinc.com,  and abhi3.pandey@samsung.com}

\begin{document}

\maketitle
\begin{abstract}
	This paper presents a novel deep learning architecture for acoustic model in the context of Automatic Speech Recognition (ASR), termed as MixNet.
	Besides the conventional layers, such as fully connected layers in DNN-HMM and memory cells in LSTM-HMM, the model uses two additional layers based on Mixture of Experts (MoE).
	The first MoE layer operating at the input is based on pre-defined broad phonetic classes and the second layer operating at the penultimate layer is based on automatically learned acoustic classes.
	In natural speech, overlap in distribution across different acoustic classes is inevitable, which leads to inter-class mis-classification.
	The ASR accuracy is expected to improve if the conventional architecture of acoustic model is modified to make them more suitable to account for such overlaps.
	MixNet is developed keeping this in mind.
	Analysis conducted by means of scatter diagram verifies that MoE indeed improves the separation between classes that translates to better ASR accuracy.
	Experiments are conducted on a large vocabulary ASR task which show that the proposed architecture provides 13.6\% and 10.0\% relative reduction in word error rates compared to the conventional models, namely, DNN and LSTM respectively, trained using sMBR criteria.
	In comparison to an existing method developed for phone-classification (by Eigen et al), our proposed method yields a significant improvement.
	
	
\end{abstract}



\section{Introduction}

%
In recent years, replacing Gaussian mixture model (GMM) hidden Markov model (HMM) \cite{hmm}, deep neural networks (DNN) \cite{dnn1, dnn2}, and more recently recurrent neural network based acoustic models have become the state-of-the-art. 
At the same time, the striving demand for human-machine interaction has called for development of new architectures for acoustic model that yields even lower word error rates (WERs) than the conventional models.
The ASR task is complicated by the fact that speech signal is a highly stochastic process in nature -- the stochasticity arises due to different speech attributes, namely, the inherent phonetic variability as well as extrinsic factors such as speaker and background noises.
In natural speech, different acoustic-classes (such as phonemes) are characterized by different frequency responses that ASR systems exploit for classification. For instance the frequency response of a voiced sounds may be very different from fricatives. However due to co-articulation and other natural variations in human speech production, the distributions of these classes may be strongly overlapped in the acoustic-space, that leads to inter-class confusion and mis-classification.
The ASR accuracy is likely to improve by modifying existing architecture of acoustic models to make them more suitable to segregate the classes, which is the objective of this paper.

Mixture of Experts (MoE) are popular in the areas of machine learning for region-dependent processing of the features using an ensemble of experts, which could be either classifiers or regressors \cite{moe, moe2, moe3, bishop}. 
Different experts are specialized to operate on specific regions in the input space.
Outputs of the experts are linearly combined using data dependent weights generated by an additional auxiliary classifier. 
The role of this classifier is to ``select" (soft or hard) the best expert that is akin to the location of the feature vector in the input space.
In this paper, we explore MoE to extend two popular deep learning architectures for acoustic modeling, namely DNN-HMM and long short term memory HMM (LSTM-HMM) \cite{lstm1}.
For brevity we will refer to this architecture as MixNet.
It is argued that the proposed architecture helps to reduce the overlap between different broad acoustic-classes, as a consequence ability of acoustic model to discriminate the classes is improved, which leads to significant improvement in ASR accuracy.
All results reported in this paper are based on DNN-HMM and LSTM-HMM models.
The main contribution of this paper is to show that MixNet outperforms these models by a significant margin, i.e., 13.6\% and 10.0\% relative reduction in WER, respectively. 

Detailed analysis is conducted to examine the distribution of the classes in the MoE space by plotting the scatter diagram. 
It is found that compared to the baseline features, MoEs provide a significant improvement in the inter-class separation, which is reflected in the improved ASR accuracy.

The organization of the rest of the paper is as follows.
In Section~\ref{related_work}, we present previous work related to proposed method.
Section~\ref{section_moe} presents the proposed MixNet architecture. 
Experimental setup and results are discussed in Section~\ref{section_expt} and  \ref{section_results}.
Section~\ref{section_conclusion} presents the conclusions.

\section{Related work}\label{related_work}
Application of MoEs in the field of speech processing is rather limited, especially in the frame-work of deep learning for ASR acoustic modeling.
One of the early work using MoE is by Jacob et al. \cite{jmoe}, who applied mixture of experts for multi-speaker vowel classification task. 
Another related work is the Hierarchical Mixture of Experts \cite{hmoe}, which learns a hierarchy of gating networks in a tree structure.
Moreover MoEs are also explored for language modeling~\cite{lmoe}.
Besides, deep mixture of experts neural-network architecture has also been explored in the field of speech enhancement \cite{dmoe},
where dedicated DNNs are used as experts in order to enhance speech for different phoneme classes given the log-spectrum of noisy speech. 
An additional gating DNN is trained to predict the class-conditional probabilities that directs a speech frame to respective experts. 
Similarly, deep recurrent mixture of experts is introduced for speech enhancement in \cite{drmoe}, where experts are deep recurrent networks. 
In an early work, MoE has been explored for ASR task in the frame-work of GMM-HMM \cite{gmoe}.

We use the architecture proposed by Eigen et al. \cite{emoe} (henceforth referred to as Eigen-DMOE) as one of our baselines in this paper. 
Extending the standard MoE architecture, in Eigen-DMOE two sets of experts ($f_{i}^{1}$,$f_{i}^{2}$) and gating networks ($g^{1}$,$g^{2}$) are introduced along with a final linear layer $f^{3}$. The final output, $\mathbf{F}(\mathbf{x})$, is produced by applying a softmax layer:
\begin{align}
\mathbf{z}^{1} &= \sum_{i=1}^{N}  {g}_{i}^{1}(\mathbf{x}) f_{i}^{1}(\mathbf{x}) \label{eq_emoe1}\\
\mathbf{z}^{2} &= \sum_{j=1}^{M}  {g}_{j}^{2}(\mathbf{z}^{1}) f_{j}^{2}(\mathbf{z}^{1}) \label{eq_emoe2} \\
\mathbf{F}(\mathbf{x})  &= \mathbf{z}^{3} = {softmax}(f^{3}(\mathbf{z}^{2})) \label{eq_emoe3}
\end{align}
In the above equations, $N$ and $M$ are number of experts in first and second MoE layers respectively.
Our work differs from Eigen-DMOE in two significant ways:
\begin{itemize}
	\item The main contribution of this paper is to evaluate MoE at input layer where it operates directly on the acoustic features.
	In addition, a second MoE layer is applied at the output, i.e just before the softmax classifier.
	The first MoE layer is interpretable and supervised, whereas the second layer is non-interpretable and unsupervised.
	In Eigen-DMOE the MoE layers are applied only at the output and are non-interpretable
	\item In original Eigen-DMOE paper, MoE based system was investigated only for mono-phone classification.
	In this paper we evaluate the proposed MixNet for large vocabulary continuous speech recognition (LVCSR) task.
	In addition, we have extended Eigen-DMOE for LVCSR.
	It is observed that MixNet outperforms it by a large margin. 
\end{itemize}


\section{Mixture of Experts Deep Neural Network}\label{section_moe}
\begin{figure}
	\centering
	\includegraphics[width=8cm]{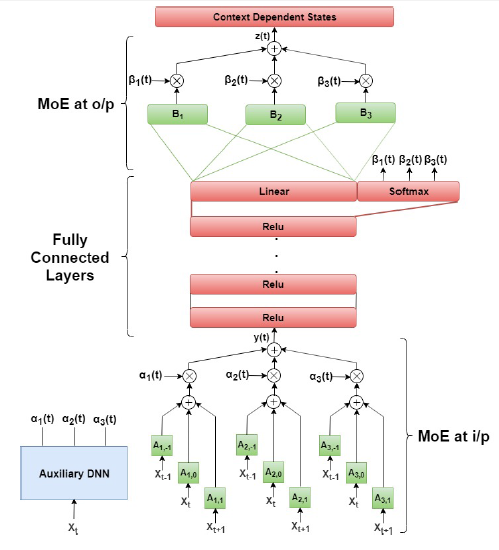}
	\caption{MixNet: (shown for 3-class MoE at input and 3-class MoE at output)}
	\label{fig:moe}
\end{figure}

The block diagram of the proposed MixNet architecture is illustrated in Figure~\ref{fig:moe}.
As shown, two MoE layers are applied to turn the standard DNN-HMM (or LSTM-HMM) acoustic models into MixNet -- the first one before the first DNN layer and the second one before the final softmax layer.
Although shown and explained only for DNN-HMM, the same developments are also applicable to LSTM-HMM.

The ``experts" in the first MoE layer, shown as the $\mathbf{A}$ matrices in green shades, consist of multiple affine transforms operating separately on different acoustic regions.
Note that this is not the standard MoE architecture as we also take temporal context ($\pm K$) into account.
We found slight improvement by using contextual MoE over standard MoE. These experts are controlled by ``gating signals" generated by an auxiliary DNN, which are used as weights for the experts. 
The auxiliary DNN is trained to classify acoustic feature vectors into broad phonetic classes.
Mathematically,
\begin{align}
\mathbf{y}(t) = \sum_{i=1}^{C_a}  \alpha_i(t) \sum_{j=-K}^{K} \left( \mathbf{A}_{ij} \mathbf{x}(t+j) + \mathbf{b}_{ij} \right)
\label{eq_moe}
\end{align}
where $C_a$ defines the number of phonetic classes, $\mathbf{x}(t)$ and $\mathbf{y}(t)$ are the input and output of the MoE layer respectively and $K$ is set to $1$.
$\alpha_i(t)$ is the posterior probability (or the gating signal) of class $i$ given the frame at time $t$, which is generated by a pre-trained DNN classifier
and the classes are (interpretable) broad phonetic classes defined in Table~\ref{3_classes_table}.

On the other hand, in the case of second MoE layer, the classes are (non-interpretable) learned in an unsupervised manner by having an additional softmax layer attached to the last hidden layer.
Note that the classifier at the second MoE is trained jointly while the classifier at the first layer is pre-trained  and training the model jointly.
The output of the softmax layers are used as gating signal for the experts denoted by the $\mathbf{B}$ matrices in Figure~\ref{fig:moe}.
The configuration gives us the freedom to choose the number classes without constraints.
Mathematically,
\begin{table}[t]
	\caption{Three Broad Phonetic Classes used in first MoE layer}
	\centering
	\begin{tabular}{|l|c|}\hline
		\multicolumn{1}{|c|}{Broad Phonetic Class} & \multicolumn{1}{|c|}{Phones} \\\hline\hline
		\multirow{3}{*}{Voiced}        &   $ae, eh, ih, iy, ix,aa$, \\
		&   $uh, ah, er, ao, ax, uw$, \\
		&   $ey, ay, aw, ow, ay$     \\\hline
		\multirow{4}{*}{Un-voiced}     &   $l, r, w, y, en, n, m, ng$,               \\
		&   $dh, jh, ts, v, z, ch, f, hh$,               \\
		&   $ch, sh, s, th, dd, kd, pd$, \\
		&   $td, k, p, t$ \\\hline
		\multirow{1}{*}{Silence}     &   $sil$               \\\hline
	\end{tabular}	
	\label{3_classes_table}
\end{table} 
\begin{align}
\mathbf{z}(t) = \sum_{i=1}^{C_b}  \beta_i(t) \left( \mathbf{B}_{i} \mathbf{y}^{L-1}(t) + \mathbf{b}_{i} \right)
\label{eq_moe1}
\end{align}
where $C_b$ defines the number of classes, $\mathbf{y}^{L-1}(t)$ and $\mathbf{z}(t)$ are the output of the ${(L-1)}^{th}$ layer and output of the second MoE.
$\beta_i(t)$ is the gating signal associated with the second MoE layer.
Temporal context was not included here.
Linear activations were used at the last hidden layer that provides input to the second MoE layer.
We also experimented with having an MoE layer after each of the hidden layers, but there was no benefit, so the results are not shown for this case.

We note that essentially the gating signal convey information about the position of the feature vectors (or the hidden layer activations in case of second MoE layer) 
in the acoustic space.
Since the above equations establish an one-to-one correspondence between the output targets of the classifier and the affine transforms,
these transforms will learn to serve as region-dependent transforms, and as a consequence, will have the flexibility to move the feature vectors belonging to different regions in different directions. 
We conjecture that the MoE layers will effectively learn to transform the features such that the overlaps between acoustic regions are reduced.
Detailed analysis that corroborates our assumption is presented later in this paper.
There are two possible choices for the posterior calculation -- hard or soft -- in this paper we focus only on the later choice as it is more optimal.
The MoE layers are trained jointly together with other layers.
After convergence the MoE parameters are frozen and the composite model is used for decoding.

\begin{table}[t]
	\caption{MixNet architectures used in experiments. Dimensions of experts are shown for DNN-HMM}
	\centering
	\begin{tabular}{|l||c|c|}\hline
		System                          &  First MoE              &   Second MoE            \\\hline\hline
		\multirow{2}{*}{MixNet-I}       &                         &  \multirow{2}{*}{-}  \\
		&                         &                      \\\cline{1-1}\cline{3-3}
		\multirow{2}{*}{MixNet-II}      &                         & 3-class \\
		&  3 phonetic class       & low-rank experts (1024$\times$512)          \\\cline{1-1}\cline{3-3}
		\multirow{2}{*}{MixNet-III}     &  full-rank experts      & 5-class  \\
		&    (143$\times$143)     & low-rank experts (1024$\times$512)                           \\\cline{1-1}\cline{3-3}
		\multirow{3}{*}{MixNet-IV}      &                         & 5-class  \\
		&                         & full-rank (1024$\times$1024) \\
		&                         & and 15-diagonal experts \\\hline
	\end{tabular}
	\label{model_table}
\end{table}

\section{Experimental Setup}\label{section_expt}
To assess the performance of the proposed MixNet, experiments are conducted on an Indian English LVCSR task.
The training set consists of approximately 500 hours of audio, 
containing about 0.5 million utterances from 2189 speakers. 
The test set consisted of 4.2 hours of audio, containing 4000 utterances from 1307 speakers. The frame accuracy is obtained on 2.3 hours  of cross validation set, containing 1870 audios from 631 speakers.
A lexicon of $2$ million words was used.
For decoding a 3-gram language model (LM) was used.
All experiments are conducted using Kaldi speech recognition toolkit \cite{kaldi}. The experimental set-up is similar to our earlier work \cite{ankitis2018}.
The GMM-HMM was trained on Mel frequency cepstral coefficients (MFCC) using LDA+MLLT \cite{mllt,stc} transformation.
There were $4749$ senones in total.
Afterwards discriminative training was applied using boosted MMI \cite{bmmi}.
This model was used to generate alignments for the DNN-HMM or LSTM-HMM model training.

\subsection{Model Arachitecture}
To compare the performance of the proposed MixNet architecture, we build the conventional LSTM-HMM and DNN-HMM systems.
We also contrast the performance with the architecture proposed by Eigen et al (Eigen-DMoE).
The input to DNN-HMM comprised $169$ dimensional features generated by mean-normalized MFCCs with context expansion of $\pm 6$, followed by LDA normalization\footnote{
	In ASR, LDA is applied for decorrelation and to improve inter-class discrimination, where the
	classes are context-dependent HMM states. It is used as the default feature
	normalization module in Kaldi's nnet3 recipes. 
	Mainly there are two short-comings of LDA. 
	Firstly, it is a linear transformation (not based on neural network) and the objective function
	is relatively weak (not cross-entropy or similar objective) which leaves scope for further improvement. 
	Secondly, in LDA a single transformation is applied irrespective of the class to which a feature vector belongs. 
	On the other hand, MixNet is a region-dependent transformation -- a separate expert operates on the feature vector depending on the class to which it belongs. 
	In the experiments LDA normalization is applied as a pre-processing module to both the baseline (DNN/LSTM and Eigen-DMoE)
	systems as well as to MixNet architectures. 
	We emphasize that MixNet is not an alternative for LDA, but it is a novel neural network architecture trained
	on top of LDA features.}.
The number of hidden units per layer was 1024 and the number of layers were tuned to get the best accuracy, i.e., 8 hidden layers.
Afterwards, Eigen-DMoE and MixNet DNN-HMMs were built using the same number of conventional hidden layers (plus MoE layers).
Similar to the original work, the Eigen-DMoE model contained 9 experts in the MoE layers. 
Besides for fair comparison, we increased the number of nodes per layer from 128 to 512 (same as number of nodes in second MoE layer of MixNet-IV) nodes in both MoE layers.
All network parameters were randomly initialized, which was followed by fine-tunning using cross-entropy loss.
Afterwards, sMBR~\cite{sMBR} training was applied to the baseline and MixNet models.

Similar to DNN-HMM, the number of hidden layers in the baseline valina LSTM-HMM was tuned to get the best accuracy (5-layers).
Each layers contained 1024 memory cells and a recurrent projection layer of size $256$.
Afterwards Eigen-DMoE and MixNet LSTMs were trained with the same number of conventional hidden layers (plus MoE layers).
The input to model comprised $65$ dimensional features ($\mathbf{x}(t)$ in Eq.~\ref{eq_moe}), obtained by splicing the mean-normalized MFCCs using a context of $\pm3$, further normalized by LDA transformation without dimensionality reduction.
40 frames are used as left context to predict the output label for the current frame. 
Output HMM state label is delayed by 5 time steps to see the information from future frames while predicting for the current frame \cite{lstm2}. 
Truncated back propagation through time (BPTT) \cite{bptt} with fixed truncation length of $20$ is used for network training.
sMBR training was applied to these models.

The other models were MixNet systems, either LSTM-HMM or DNN-HMM, defined in Table~\ref{model_table}.
The model architectures were same as baseline, albeit the additional MoE layers as shown in the Figure~\ref{fig:moe}.
The first MoE layer consisted of 3 broad phonetic classes defined in Table~\ref{3_classes_table}.
We experimented with 2 (speech and non-speech) and 13 phonetic classes, but 3-class yielded the optimal result.
The number of classes in the second MoE layer were varied in the experiments.
MixNet-I uses an MoE layer only at the input, but no second MoE.
We note here that the second MoE layer affine transforms are connected to the output layer, so they can potentially consume a large number of parameters.
To verify that the observed performance improvement is not just because of having more parameters,
we make sure the number of parameters in the MixNet models are approximately the same (or less) as the baseline models.
To achieve this we used either low rank experts \cite{lowrank} or full-rank but diagonal experts in the second MoE layer.
Specifically, MixNet-II and III use low-rank matrices of dimension 512, whereas MixNet-IV uses multi-diagonal matrices with 15 diagonal entries on each side of the principal diagonal.
The training algorithms and input features were same as the baseline systems, except that for fair comparison the input context was reduced to $\pm 5$ in case of MixNet DNN-HMMs.
This is to ensure that the overall context seen by the baseline DNN-HMM and MixNet are same (note that the first MoE layer includes a context of $\pm 1$).
Similarly, the context was reduced from $3$ to $2$ in the case of MixNet LSTM-HMM.

The auxiliary classifier in both DNN and LSTM MixNet is a three layer DNN network each containing 512 ReLU units.
The model was trained to predict the posterior probabilities of broad phonetic classes defined in Table~\ref{3_classes_table}.
The output layer consisted of a soft-max layer containing 3 units corresponding to the broad phonetic classes.
The same 500 hours of training set was used for the training.
The network was trained using cross-entropy loss.
The frame accuracy of auxiliary classifier on a cross-validation set was 94\%. 

\section{Results}\label{section_results}

\begin{figure}[t]
	\centering
	\includegraphics[width=8.8cm]{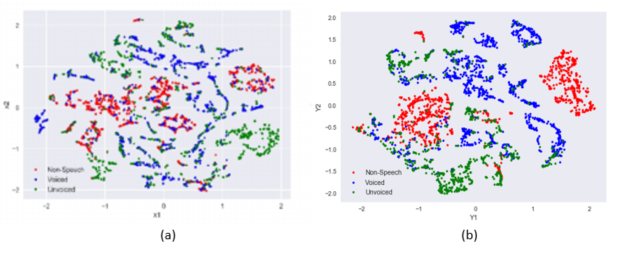}
	\caption{t-SNE plots comparing the baseline features (left) and the {\em first MoE layer} output (right). It is noted that in MoE space, the classes are better separated than the baseline features, especially speech and non-speech classes.}
	\label{fig:scatter_diagram_before}
\end{figure}

First we present the analysis that shows that our assumption described earlier is valid. All analysis is carried out for DNN-HMM.
We used the t-SNE \cite{tsne},  to map the distributions of baseline features (i.e., $\mathbf{x}(t)$ in Eq.~\ref{eq_moe}) and features at the output of the first MoE layers (i.e., $\mathbf{y}(t)$ in Eq.~\ref{eq_moe}).
Figure~\ref{fig:scatter_diagram_before} shows the t-SNE for $10$ utterances randomly selected from the training set.
The 3-classes (Table~\ref{3_classes_table}) are marked in different colors.
It may be noted that MoE features are more discriminative than the baseline features, in particular, the non-speech class is well separated from the other two classes.

Percentage of WERs (\%WER) are obtained on test set and frame accuracy on cross-validation set, comparing the MoE systems with the baseline, are presented in Table~\ref{wer_table} and \ref{wer_table1} for DNN-HMM and LSTM-HMM respectively.
The general observation from Table~\ref{wer_table} is that all MixNet DNN-HMM systems outperform the baselines DNN-HMM and Eigen-DMoE by a considerable margin.
MixNet-I gives an improvement of 1.3\% absolute compared to  DNN-HMM (8 fully connected layers), 
whereas MixNet-II, consisting of 3 classes in the second MoE layer and low rank experts, gives no additional improvement. 
This may be because the expected improvement is nullified by the loss incurred due to low-rank projection.
We note that this model has about 1.7 million parameters less than MixNet-I due to the low-rank layer.
MixNet-III, comprising of 5 classes in the second MoE layer, gives a lower WER than MixNet-I and MixNet-II, which amounts to an improvement of 12.1\% relative compared to the DNN-HMM baseline.
The best performance is observed with MixNet-IV. It gives a further improvement, which amounts to 13.4\% (2.1\% absolute) and 6.2\% (0.9\% absolute) relative compared to the baseline DNN-HMM (8 FC layers) and Eigen-DMoE, respectively. 
Results also translates to sequence discriminative training which amount for 13.6\% (2.0\% absolute) and 7.3\% (1.0\% absolute) relative improvement compared to DNN-HMM (8 FC layers) and Eigen-DMoE, respectively.
It is noted that in the special case when $C_a$ and $C_b$ are 1, MixNet reduces to conventional baseline with two additional fully connected layers, in which case the performance is reduced to baseline.

The improvements also carry over to LSTM-HMM models. 
We note from Table~\ref{wer_table1} that MixNet-IV yields 1.4\% and 1.1\% absolute gains compared to LSTM-HMM and Eigen-DMoE LSTM-HMM, respectively, which equal to 9.4\% and 7.5\% relative. The Improvement translates to sequence discriminative training as well.
MixNet-IV gives an absolute improvement of 1.4\% and 1.0\% compared to LSTM-HMM and Eigen-DMoE.

\begin{table}[t]
	\caption{Results on DNN-HMM. FC refers to fully connected layer. Frame Accuracies are shown on CV Set while WERs are shown on Test Set.}
	\centering
	\begin{tabular}{l|c|c}\hline
		\multirow{2}{*}{Architecture}                     & \% Frame &  \multirow{2}{*}{(\%) WER}               \\
		& Accuracy &                 \\\hline\hline 
		DNN - 6 FC layers                & 59.1       & 16.6   \\ \
		DNN - 7 FC layers                & 60.7      & 16.0   \\ \
		DNN - 8 FC layers                & 61.3     & 15.7   \\ \
		DNN - 9 FC layers                & 61.5      & 15.7   \\ \
		DNN - 8 FC layers $+$ sMBR       &   -       & 14.7   \\\hline \
		Eigen - DMoE - 8 FC  + 2 MoE              &  62.1         & 14.5   \\ \
		Eigen - DMoE - 8 FC + 2 MoE $+$ sMBR      &  -         & 13.7
		\\\hline \  
		MixNet-I - 8 FC + 1 MoE                     &  62.8         & 14.4 \\ \
		MixNet-II - 8 FC + 2 MoE                    & 61.9          & 14.5  \\ \
		MixNet-III - 8 FC + 2 MoE                   & 63.2         & 13.8   \\ \
		MixNet-IV - 8 FC + 2 MoE ($C_a$=$C_b$=$1$ )                    &  61.6        & 15.6 \\ \
		MixNet-IV - 8 FC + 2 MoE                    &  63.4        & 13.6 \\ \
		MixNet-IV - 8 FC + 2 MoE + sMBR             &  -         & 12.7
		\\\hline       
	\end{tabular}
	\label{wer_table}
\end{table}

\begin{table}[t]
	\caption{Results on LSTM-HMM. Frame Accuracies are shown on CV Set while WERs are shown on Test Set.}
	\centering
	\begin{tabular}{l|c|c}\hline
		\multirow{2}{*}{Architecture}                              & \% Frame    &  \multirow{2}{*}{(\%) WER}               \\
		& Accuracy   &                 \\\hline\hline
		LSTMP - 4 LSTM layers                     &  61.9        & 15.4   \\\
		LSTMP - 5 LSTM layers                     &  62.1        & 14.9   \\\
		LSTMP - 6 LSTM layers                     &  62.4        & 14.9   \\\
		LSTMP - 5 LSTM layers $+$ sMBR            &  -         & 14.0 \\\hline \
		Eigen - DMoE - 5 LSTM + 2 MoE             &  62.8          & 14.6   \\ \
		Eigen - DMoE - 5 LSTM + 2 MoE + sMBR      &  -         & 13.6   \\\hline \ 
		MixNet-IV - 5 LSTM  + 2 MoE                  &  63.3        & 13.5 \\\
		MixNet-IV - 5 LSTM  + 2 MoE  + sMBR          &  -         & 12.6
		\\\hline       
	\end{tabular}
	\label{wer_table1}
\end{table}

\section{Conclusions}\label{section_conclusion}
This paper investigates a novel architecture for acoustic modeling using Mixture of Experts in the context of automatic speech recognition.
Experiments conducted on a large vocabulary ASR task show that MoE helps to reduce the overlap in distributions of different acoustic classes, which leads to significant reduction in WER compared to strong baselines, namely, DNN and LSTM.
The scatter diagram verifies that the MoE features indeed have higher inter-class separation than the baseline features, which is translated to lower WER compared to baselines.

Future direction includes 
noise robustness of MixNet in far-field scenario.
It will also be interesting to investigate the method in case of lower frame rate (LFR) \cite{lfr} acoustic models.

\newpage
\cleardoublepage





\end{document}